\documentclass[aps,prb,twocolumn,showpacs,groupedaddress]{revtex4-1}
\usepackage{amsfonts}
\usepackage{amsmath}
\usepackage{amssymb}
\usepackage{graphicx}

\begin{document}

\title[Negative dc conductivity]{Cyclotron-resonance-induced negative dc conductivity in a
two-dimensional electron system on liquid helium}
\author{Yu.P. Monarkha}
\affiliation{Institute for Low Temperature Physics and Engineering, 47 Lenin Avenue, 61103
Kharkov, Ukraine}

\begin{abstract}
We theoretically predict instability of a zero-dc-current state
of the two-dimensional electron system formed on the surface of
liquid helium induced by the cyclotron resonance (CR).
This conclusion follows from the theoretical analysis of the
dc magnetoconductivity which takes into account the contribution
from radiation in an exact way. A many-electron model
of the dynamic structure factor of the 2D Coulomb liquid is used
to describe the influence of strong internal forces acting between electrons.
For low electron densities and high amplitudes of the microwave field,
the dc magnetoconductivity is shown to become negative in the vicinity
of the CR which causes the instability. This effect is strongly
suppressed by Coulomb forces in the region of high densities.
\end{abstract}

\pacs{73.40.-c,73.20.-r,73.25.+i, 78.70.Gq}



\maketitle

Microwave-induced resistance oscillations and zero-resistance states (ZRS)
observed in a two-dimensional (2D) electron gas subjected to a transverse
magnetic field~\cite{ZudSim-2001,ManSme-2002,ZudDu-2003}
represent a surprising discovery in condensed
matter physics. A number of theoretical mechanisms have been proposed to
explain these oscillations and
ZRS~\cite{DurSac-2003,RyzSur-2003,DmiVav-2005,Mik-2011,CheShe-2009}.
Still, by now the origin of the phenomenon is controversial.
A crucial result~\cite{AndAleMil-2003},
independent of the details of the microscopic mechanism, is that the ZRS can
be explained by an assumption that the longitudinal linear response
conductivity $\sigma _{xx}$ is negative in appropriate ranges of the
magnetic field $B$. In this case, a zero-current state is unstable and the
system spontaneously develops a nonvanishing local current density.

The above noted phenomena were observed in high quality GaAs/AlGaAs
heterostructures. There is another extremely clean 2D electron system formed
on the surface of liquid helium which exhibits remarkable quantum
magnetotransport phenomena~\cite{MonKon-book-2004} although
it is a nondegenerate system of strongly interacting electrons. Since
electron gas degeneracy is not a crucial point for some of the theoretical
mechanisms explaining ZRS in semiconductor systems, these mechanisms
potentially can be applied to surface electrons (SEs) on liquid helium as
well. Complementary studies of these phenomena in the system of SEs on
liquid helium could help with identification of the origin of ZRS observed
in GaAs/AlGaAs heterostructures.

It should be noted that another kind of magnetoconductivity oscillations and
ZRS (zero-$\sigma _{xx}$ states to be exact) was already observed in the
system of SEs on liquid helium~\cite{KonKon-2009,KonKon-2010} when the energy
of excitation of the second surface subband ($\Delta _{2}-\Delta _{1}$) was
tuned to the resonance with the microwave (MW) frequency.
These phenomena were explained~\cite{Mon-2011} by nonequilibrium population of the second surface
subband which triggers quasi-elastic intersubband decay processes
accompanying by electron scattering against or along the driving force,
depending on the ratio $\left( \Delta _{2}-\Delta _{1}\right) /\hbar \omega
_{c}$ (here $\omega _{c}$ is the cyclotron frequency).
An important evidence for identification of the mechanism of oscillations and
ZRS observed for SEs on liquid helium was recently found in studies of the
Coulombic effect on positions of conductivity extrema~\cite{KonMonKon-2013}.

The most frequently discussed mechanism of negative conductivity effects
called the displacement mechanism was proposed already in 1969 by
Ryzhii~\cite{Ryz-1969}. In this model, an
electron scattering by an impurity potential
accompanying by absorption of a photon is the origin of the negative linear
response conductivity. In recent developments of the model~\cite{Par-2004,LyaPat-2006},
the contribution from radiation is taken into account exactly
which is important for high MW powers. Even though, in semiconductor systems,
other mechanisms reportedly~\cite{DmiMirPol-2012,Mik-2014} could give a stronger effect, it is
very attractive to study the displacement model for the 2D electron system
on liquid helium.

In the displacement model, the strength of the effect depends on
the product of two dimensionless parameters:
\begin{equation}
\text{\ $\lambda $}=\text{$\frac{eE_{\mathrm{ac}}^{(0)}l_{B}}{\hbar \omega }$%
},\text{ \ \ }\varkappa =\frac{\omega _{c}^{2}}{\left\vert \omega
^{2}-\omega _{c}^{2}\right\vert } , \label{e1}
\end{equation}%
where $E_{\mathrm{ac}}^{(0)}$ is the amplitude of the MW field, and $l_{B}=%
\sqrt{\hbar c/eB}$ is the magnetic length. For a fixed ratio $\omega /\omega
_{c}$, the value of $\varkappa $ is the same in both semiconductor and SE
systems. The effective mass of SEs is very close to the free electron
mass $m_{e}$, while the effective mass of semiconductor electrons is much
smaller: $m_{e}^{\ast }\simeq 0.064\,m_{e}$. Therefore, at fixed $E_{\mathrm{ac%
}}^{(0)}$ and $\omega $, in experiments with SEs on liquid helium
$\lambda $ is smaller than it is for semiconductor electrons by
the factor $\sqrt{m_{e}^{\ast }/m_{e}}\simeq 1/4$. The reduction of $\lambda $
can be well compensated by approaching the cyclotron resonance (CR)
condition $\omega _{c}\rightarrow \omega $, which increases
$\varkappa $. Thus, for SEs on liquid helium, negative dc conductivity
(similar to that of the semiconductor model) could be expected in the vicinity
of the CR. Since the average Coulomb interaction energy per SE $U_{C}$
is usually much larger than the average kinetic energy, a many-electron
treatment of the displacement model is required.

In this work, we report results of a theory of the displacement mechanism of
negative dc conductivity applied to SEs on liquid helium interacting with
capillary-wave quanta (ripplons). The ac electric field is taken into account in an
exact way similar to Refs.~\onlinecite{Par-2004,LyaPat-2006}. Scattering with
ripplons is described using a perturbation theory. Strong Coulomb
interaction is taken into account employing a model based on the dynamic
structure factor (DSF) of a 2D electron liquid. We found that results of the
many-electron treatment drastically depend on SE density $n_{s}$, which
allows us to predict the range of $n_{s}$ where negative conductivity effects
can be observed.

We will use the Landau gauge in which a momentum in $y$ direction ($p_{y}$)
is a good quantum number. The dc and ac electric fields ($\mathbf{E}_{%
\mathrm{dc}}$ and $\mathbf{E}_{\mathrm{ac}}$) are taken to be parallel to
the $x$-axis. The ac field $E_{\mathrm{ac}}\left( t\right) =E_{\mathrm{ac}%
}^{\left( 0\right) }\cos \omega t$. In the absence of scatterers, the exact
solution of the single-electron Hamiltonian can be written
as~\cite{Hus-53,Par-2004}
\[
\text{\ }\psi _{n,X}\left( x,y,t\right) =e^{i\vartheta \left( x,y,t\right)
}\exp \left\{ i\frac{X}{l_{B}}\beta \sin \omega t\right\} \text{ }\times
\text{\ }
\]%
\begin{equation}
\times \exp \left( -iXy/l_{B}^{2}\right) \varphi _{n}\left( x-X-\xi \left( t\right)
,t\right) ,  \label{e2}
\end{equation}%
where

\begin{equation}
\text{\ \ \ \ }X=-\frac{cp_{y}}{eB}-\frac{eE_{\mathrm{dc}}}{m_{e}\omega
_{c}^{2}},\text{ \ \ \ }\beta =\lambda \frac{\omega _{c}^{2}}{\left( \omega
^{2}-\omega _{c}^{2}\right) },\text{ \ }  \label{e3}
\end{equation}%
$\varphi _{n}\left( x,t\right) $ is the well-known solution for the unforced
quantum harmonic oscillator, $n$ is the Landau level index, and $\xi \left(
t\right) $ is the classical solution of the forced harmonic oscillator, $\xi
=eE_{\mathrm{ac}}^{(0)}\cos \omega t/m\left( \omega ^{2}-\omega _{c
}^{2}\right) $. The resonant denominator of $\beta $ originates from $\dot{\xi }$.
The exact expression for $\vartheta \left( x,y,t\right) $ is not important
in the following treatment.

The interaction with ripplons causes electron scattering between different
states given in Eq.~(\ref{e2}). The dc conductivity $\sigma _{xx}$ of SEs can
be found from the equation for current density $j_{x}=-en_{s}l_{B}^{2}\sum_{\mathbf{q}%
}q_{y}\bar{w}_{\mathbf{q}}$, where $\bar{w}_{\mathbf{q}}$ is the average
probability of electron scattering with the momentum exchange $\hbar \mathbf{%
q}$, and $l_{B}^{2}q_{y}$ represents a change of the orbit center number $X$
for such a process. The effective collision frequency $\nu _{\mathrm{eff}}$,
entering the usual conductivity form, is found as%
\begin{equation}
\nu _{\mathrm{eff}}=-\frac{1}{m_{e}V_{H}}\sum_{\mathbf{q}}\hbar q_{y}\bar{w}%
_{\mathbf{q}}\left( V_{H}\right) ,  \label{e4}
\end{equation}%
where $V_{H}=cE_{\mathrm{dc}}/B$\ is the absolute value of the
Hall velocity.
The $\bar{w}_{\mathbf{q}}$ depends on $E_{\mathrm{dc}%
}$ because, in addition to $\varepsilon _{n}=\hbar \omega _{c}\left(
n+1/2\right) $, we have a term $eE_{\mathrm{dc}}X$. For
nondegenerate electrons, the probability $w_{\mathbf{q}}$ is independent of
the quantum number $X$ due to $eE_{\mathrm{dc}}\left( X^{\prime }-X\right) =\hbar
q_{y}V_{H}$. In this case, $w_{\mathbf{q}}$ can be averaged over Landau
level numbers only, assuming an equilibrium distribution $Z_{\Vert
}^{-1}\exp \left( -\varepsilon _{n}/T_{e}\right) $ (here
$Z_{\Vert} $ is the partition function).

In a single-electron treatment, $\bar{w}_{\mathbf{q}}$ can
be found in terms of the DSF of a nondegenerate 2D electron gas
\[
S\left( q,\Omega \right) =\frac{2}{\pi \hbar Z_{\Vert }}\sum_{n,n^{\prime
}}I_{n,n^{\prime }}^{2}\left( x_{q}\right) \times
\]
\begin{equation}
 \int d\varepsilon e^{-\varepsilon
/T_{e}}g_{n}\left( \varepsilon \right) g_{n^{\prime }}\left( \varepsilon
+\hbar \Omega \right) ,  \label{e5}
\end{equation}%
where $g_{n}\left( \varepsilon \right) =-\mathrm{Im}G_{n}\left( \varepsilon
\right) $ represents the Landau level density of states, $G_{n}\left(
\varepsilon \right) $ is the single-electron Green's function, $%
x_{q}=q^{2}l_{B}^{2}/2$,
\begin{equation}
I_{n,n^{\prime }}\left( x\right) =\sqrt{\frac{\min \left( n,n^{\prime
}\right) !}{\max \left( n,n^{\prime }\right) !}}x^{\frac{%
\left\vert n^{\prime }-n\right\vert }{2}}e^{-\frac{x}{2}}
L_{\min \left( n,n^{\prime }\right) }^{\left\vert n^{\prime }-n\right\vert
}\left( x\right) ,  \label{e6}
\end{equation}%
and $L_{n}^{m}\left( x\right) $ are the associated Laguerre polynomials.
This representation is similar to that of the theory of thermal neutron (or
X-ray) scattering by solids, where the scattering cross section of a
particle flux is expressed in terms of the DSF of the target.

Comparing with the case $E_{\mathrm{ac}}^{\left( 0\right) }=0$, matrix
elements, describing electron scattering,
contain the additional factor $\exp \left( -iq_{y}l_{B}\beta \sin
\omega t\right) $.
Using the expansion $e^{iz\sin \phi
}=\sum_{m}J_{m}\left( z\right) e^{im\phi }$ [here $J_{m}\left( z\right) $ is
the Bessel function], the procedure of finding scattering probabilities can
be reduced to a quite usual treatment. Then, for ripplon
creation (index $+$) and destruction (index $-$) processes, $\bar{w}_{\mathbf{%
q}}^{(\pm )}\left( V_{H}\right) $ can be found as
\[
\bar{w}_{\mathbf{q}}^{(\pm )}\left( V_{H}\right) =
\frac{\left\vert \bar{C}_{\mathbf{q}}^{\pm
}\right\vert ^{2}}{\hbar ^{2}}\sum_{m=-\infty }^{\infty }J_{m}^{2}\left(
\beta q_{y}l_{B}\right) \times
\]
\begin{equation}
 S\left( q,-q_{y}V_{H}+m\omega \mp \omega
_{r,q}\right) ,  \label{e7}
\end{equation}%
where $\bar{C}_{q}^{\pm }=V_{r,q}Q_{q}\left[ N_{q}^{(r)}+1/2\pm 1/2\right]
^{1/2}$, $N_{q}^{(r)}$ is the ripplon distribution function, $V_{r,q}$ is the
electron-ripplon coupling~\cite{MonKon-book-2004}, $Q_{q}=\sqrt{\hbar
q/2\rho \omega _{r,q}}$, $\omega _{r,q}\simeq \sqrt{\alpha /\rho }q^{3/2}$, $%
\alpha $ and $\rho $ are the surface tension and mass density of liquid
helium respectively.

Generally, the structure of Eq.~(\ref{e7})
is similar to that found for other scattering mechanisms important for
semiconductor electrons~\cite{Ryz-2003,Par-2004,LyaPat-2006}. The main
advantage of the form of Eq.~(\ref{e7}) is that we can employ the properties of
the equilibrium DSF and model the effect of
Coulomb interaction using the DSF of strongly interacting SEs. Such a
possibility appears under the condition $l_{B}\ll a$ (here $a$ is a typical
electron spacing) which allows to consider a fluctuational electric field $%
\mathbf{E}_{\mathrm{f}}$, acting on a particular
electron, as a quasi-uniform field~\cite{DykKha-1979}. This reduces the
many-electron problem to a single-electron dynamics. Even in this limit, the
situation remains to be very complicated; still an accurate form of the DSF of
the 2D Coulomb liquid in a magnetic field can be found~\cite{MonKon-book-2004}%
:
\begin{equation}
S\left( q,\Omega \right) =\frac{2\sqrt{\pi }}{Z_{\parallel }}%
\sum_{n,n^{\prime }}\frac{I_{n,n^{\prime }}^{2}}{\gamma _{n,n^{\prime }}}%
\exp \left[ -\frac{\varepsilon _{n}}{T_{e}}-P_{n,n^{\prime }}\left( \Omega
\right) \right] \text{ },  \label{e8}
\end{equation}%
where%
\begin{equation}
P_{n,n^{\prime }}=\frac{\left[ \Omega -\left( n^{\prime }-n\right) \omega
_{c}-\phi _{n}\right] ^{2}}{\gamma _{n,n^{\prime }}^{2}},\text{ }\phi _{n}=%
\frac{\Gamma _{n}^{2}+x_{q}\Gamma _{C}^{2}}{4T_{e}\hbar },  \label{e9}
\end{equation}%
\begin{equation}
\hbar \gamma _{n,n^{\prime }}=\sqrt{\frac{\Gamma _{n}^{2}+\Gamma _{n^{\prime
}}^{2}}{2}+x_{q}\Gamma _{C}^{2}},  \label{e10}
\end{equation}%
$\Gamma _{n}$ is the collision broadening of Landau levels, $\Gamma _{C}=%
\sqrt{2}eE_{\mathrm{f}}^{(0)}l_{B}$ , and $E_{\mathrm{f}}^{(0)}\simeq 3\sqrt{T_{e}}n_{s}^{3/4}$ is
the typical fluctuational electric field~\cite{FanDykLea-1997} under
the condition $U_{C}/T>10$.

Remarkably, the DSF of the Coulomb liquid with strong interaction given in
Eq.~(\ref{e8}) is similar to the DSF of non-interacting electrons (the later
corresponds to the regime $\Gamma _{C}\ll \Gamma _{n}$). The proportionality
factor $1/\gamma _{n,n^{\prime }}$ reflects the singular nature of the magnetotransport
in 2D systems. For $\Gamma _{C}=0$, eventually, it leads to the enhancement
factor $\hbar \omega _{c}/\Gamma _{n}$
of the SCBA theory, which describes the effect of multiple electron scattering.
The fluctuational electric filed drives an electron from a scatterer which reduces
multiple scattering by increasing $\gamma _{n,n^{\prime }}$ given in Eq.~(\ref{e10}).
As the function of frequency, the DSF has maxima
near Landau excitation energies. The fluctuational field introduces an
additional broadening of these maxima $\sqrt{x_{q}}\Gamma _{C}$ and the
shift in their positions $\phi _{C}=x_{q}\Gamma _{C}^{2}/4T_{e}\hbar
$. This form of the DSF describes well the magnetotransport properties of SEs%
~\cite{MonKon-book-2004} even those induced by the intersubband MW
resonance~\cite{KonMonKon-2013}.

In most cases, the ripplon energy can be disregarded in the frequency
argument of the DSF which allows to consider electron scattering as
quasi-elastic (in this limit $\bar{C}_{q}^{\pm }\rightarrow \bar{C}_{\mathbf{%
q}}=V_{r,q}Q_{q}\sqrt{T/\hbar \omega _{r,q}}$). Then, using the property of
the equilibrium DSF, $S\left( q,-\Omega \right) =\exp \left( -\hbar \Omega
/T_{e}\right) S\left( q,\Omega \right) $, the effective collision frequency
can be represented as the sum $\nu _{\mathrm{eff}}=\sum_{m=0}^{\infty }$ $%
\nu _{m}$, where%
\begin{equation}
\nu _{0}=\frac{1}{m_{e}T_{e}}\sum_{\mathbf{q}}q_{y}^{2}\left\vert \bar{C}_{%
\mathbf{q}}\right\vert ^{2}J_{0}^{2}\left( \beta q_{y}l_{B}\right) S\left(
q,0\right) ,  \label{e11}
\end{equation}%
and%
\[
\nu _{m}=\frac{2}{m_{e}\hbar }\sum_{\mathbf{q}}q_{y}^{2}\left\vert \bar{C}_{%
\mathbf{q}}\right\vert ^{2}J_{m}^{2}\left( \beta q_{y}l_{B}\right) \times
\]%
\begin{equation}
\left[ \left( 1-e^{-\frac{m\hbar \omega }{T_{e}}}\right) S^{\prime }\left(
q,m\omega \right) +\frac{\hbar }{T_{e}}e^{-\frac{m\hbar \omega }{T_{e}}%
}S\left( q,m\omega \right) \right]   \label{e12}
\end{equation}%
for $m>0$. The derivative $S^{\prime }\equiv \partial S/\partial \Omega $
appears because of the linear expansion of the function $\bar{w}_{\mathbf{q}%
}\left( V_{H}\right) $ in Eq.~(\ref{e4}). The Eq.~(\ref{e11}) follows from the
relationship $S^{\prime }\left( q,0\right) =\left( \hbar /2T_{e}\right)
S\left( q,0\right) $. Usually, the second term in square brackets
of Eq.~(\ref{e12}) is very small. In the following, we shall consider
the regime of sharp maxima of $S\left( q,\Omega \right) $,
realized at $\gamma _{n,n^{\prime }}\ll \omega_{c}$.

Since $\lambda $, entering the definition of $%
\left\vert \beta \right\vert =\lambda \varkappa $, is usually very small,
we shall concentrate on effects induced
by the CR, when $\omega _{c}$ is quite close to $\omega $.
Sometimes we shall use a damping form $\sqrt{\left( \omega ^{2}-\omega
_{c}^{2}\right) ^{2}+4\gamma _{\ast }^{2}\omega ^{2}}$ instead of $%
\left\vert \omega ^{2}-\omega _{c}^{2}\right\vert $ in the denominator of $%
\varkappa $ given in Eq.~(\ref{e1}). This denominator originates from
the classical equation; therefore, it is reasonable to set the damping parameter $%
\gamma _{\ast }$ to its classical value $\nu _{cl}$.

The Eq.~(\ref{e11}) indicates
that $\nu _{0}$, as the function of $\omega -\omega _{c}$, has a symmetrical
minimum at $\omega _{c}=\omega $. To the contrary, the next term $\nu _{1}$
and the following terms $\nu _{m}$ with $m>1$ have an asymmetrical shape of
a derivative of a maximum affected by the symmetrical factor $J_{m}^{2}\left(
\lambda \varkappa q_{y}l_{B}\right) $. For noninteracting electrons, $S^{\prime }\left(
q,\omega \right)$ has a negative minimum at $\omega -\omega _{c}=\gamma _{0,1}/\sqrt{2}$.
At this point, the whole sum $\nu _{\mathrm{eff}}$
becomes negative already at $\lambda
=1.4\cdot 10^{-3}$ due to the proximity of the CR condition and the extreme
sharpness of Landau levels. This estimate is very promising for experimental
studies of negative conductivity effects in the system of SEs on liquid
helium.

\begin{figure}[tbp]
\begin{center}
\includegraphics[width=11.0cm]{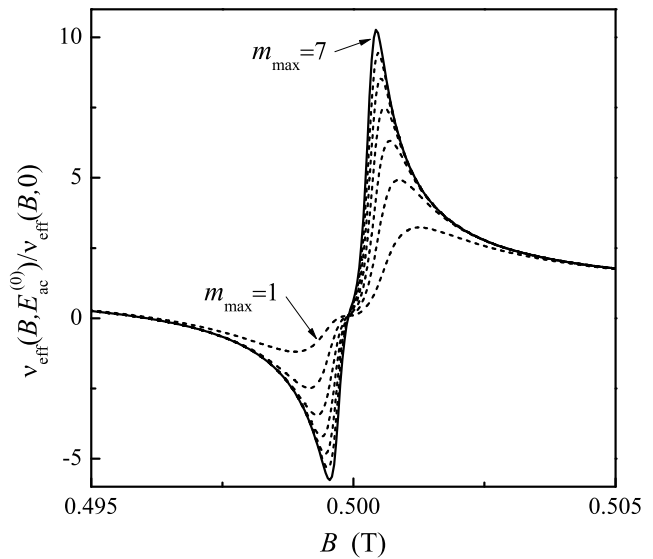}
\end{center}
\caption{ Contributions from partial sums $\sum_{m=0}^{m_{\max }}$ $\nu _{m}$
to $\nu _{\mathrm{eff}}$ normalized vs the magnetic field $B$ for
a sequence of $m_{\max }$: from $m_{\max }=1$ to $m_{\max }=7$ (solid). The conditions are the
following: $T=0.2\,\mathrm{K}$ (liquid $^{4}\mathrm{He}$), $n_{s}=10^{6}\,\mathrm{cm}^{-2}$,
and $E_{\mathrm{ac}}^{(0)}=0.05\,\mathrm{V/cm}$.
} \label{f1}
\end{figure}

At the chosen value of $\omega -\omega _{c}$, the sum over $m$ converges
quite rapidly. Still, each next term in the
sum of $\nu _{\mathrm{eff}}$ has its own minimum
which is closer to the point $\omega _{c}=\omega $.
This follows from Eqs.~(\ref{e8}) and (\ref{e9}): for each energy exchange $\hbar\Omega=\hbar m\omega $ there
is a term with $n'-n=m$ having a sharp maximum near $\omega _{c}=\omega $ (the sharpness
of the maximum increases with $m$).
The derivative of such a term contributes to $\nu _{m}$ of Eq.~(\ref{e12}).
Since the effect of $E_{\mathrm{ac}}$ eventually comes from $\dot{\xi}(t)$, an approach to
the resonance condition is equivalent to an effective increase in $E_{\mathrm{ac}}$
which explains the importance of multi-photon terms.
Thus, in the vicinity of the resonance, a substantial number of $\nu _{m}$ should be taken into
account. This situation is illustrated in Fig.~\ref{f1} where partial sums $%
\sum_{m=0}^{m_{\max }}$ $\nu _{m}$ with different $m_{\max }$ are shown as
functions of $B$. Here, the damping parameter $\gamma _{\ast }=\nu _{cl}$,
and the many-electron DSF is taken into account for
$n_{s}=1\cdot 10^{6}\mathrm{cm}^{-2}$.
It is clear that an inclusion of higher terms only enhances the effect of
negative conductivity making the minimum deeper and shifting its position
closer to the point $\omega _{c}=\omega $.

\begin{figure}[tbp]
\begin{center}
\includegraphics[width=11.0cm]{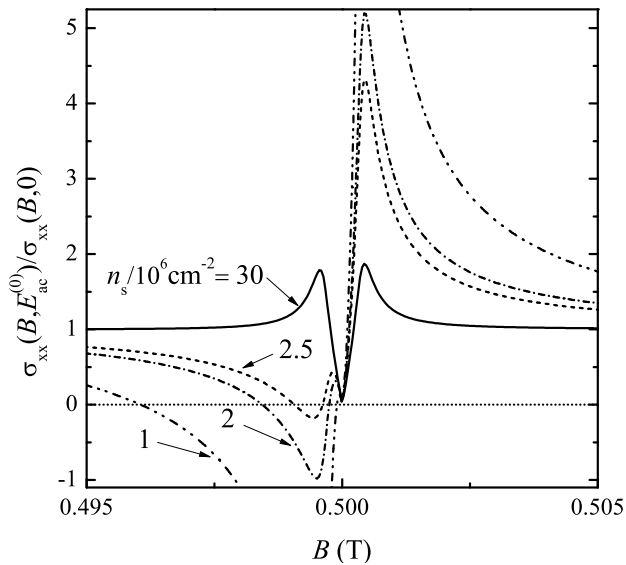}
\end{center}
\caption{ The magnetoconductivity $\sigma_{xx}$ normalized vs $B$ for different
electron densities: $n_{s}/10^{6}\,\mathrm{cm}^{-2}=1$ (dash-dot-dotted),
$2$ (dash-dotted), $2.5$ (dashed), and $30$ (solid). Here $m_{\max }=7$. Other
conditions are the same as in Fig.~\ref{f1}
.} \label{f2}
\end{figure}

The Fig.~\ref{f2} illustrates how Coulomb forces affect $\sigma _{xx}\left(
B\right) $. Here it was instructive to set $\gamma _{\ast }=0$. One can see that an
increase in $n_{s}$ strongly suppresses the conductivity minimum and maximum
near the CR without substantial changes in their positions and broadening.
This is contrary to the Coulombic effect reported previously for
intersubband displacement mechanism~\cite{Mon-2012} and for
conductivity oscillations with $\omega /\omega _{c}\geq 2$ caused by
one-photon assisted scattering~\cite{Mon-2014}.

For $n_{s}\gtrsim 2\cdot 10^{6}\,\mathrm{cm}^{-2}$, an
additional maximum is formed at $\omega -\omega
_{c}>0$. Both the
minimum and the maximum of the region $\omega -\omega _{c}>0$ are moving up
when $n_{s}$ increases. Eventually, the curve $\sigma _{xx}\left( B\right) $
obtains the shape with two maxima settled in the regions $\omega -\omega
_{c}>0$ and $\omega -\omega _{c}<0$ (the later one is higher) and with a
strong minimum positioned between the maxima at $\omega -\omega _{c}$. It
should be noted that a similar shape of the dc conductivity affected by
the CR was experimentally observed for the vapor atom scattering regime~\cite%
{PenTreShi-2000}.
An important conclusion which follows from Fig.~\ref{f2} is
that electron density should be rather small
to obtain the negative conductivity regime induced be the
CR.

Recent experiments~\cite{BadAbdKon-2014} indicate
an unusually large expansion of the electron system in a lateral direction, which
cannot be understood in the framework of the generally accepted
effective electron temperature approximation.
Electron densities used in this experiment were rather high $%
n_{s}>40\cdot 10^{6}\,\mathrm{cm}^{-2}$ which does not allow us to use directly
an explanation based on the negative dc conductivity. Still,
for electron temperatures $T_{e}$ estimated there, the system
enters the regime $U_{c}/T_{e}<1$, where the fluctuational field model fails
and the single-electron treatment could be a much better approximation. In
this regime, the negative dc conductivity could appear even for a high
$n_{s}$, because $\nu _{0}$ decreases with $T_{e}$ stronger than the
sign-changing terms $\nu _{m}$ with $m\geq 1$.
In a Corbino geometry, the formation of a steady ring current
$j$, making $\sigma _{xx}(j)=0$, should be accompanied by a lateral redistribution of SEs.
Predictions on properties of the ring current are
possible only in a nonlinear (in $E_{\mathrm{dc}}$) treatment which requires separate investigations.
Nevertheless, the present theory allows us to formulate experimental conditions,
where negative dc conductivity effects can be observed.

In summary, we have investigated theoretically the influence of cyclotron resonant excitation on
the dc magnetoconductivity of the highly correlated 2D electron system formed on the surface
of liquid helium. In the low electron density region ($n_{s}\lesssim 10^{6}\, \mathrm{cm^{-2}}$), where the
Coulomb interaction is weak enough, the dc magnetoconductivity is shown to
reach negative values for quite usual amplitudes of the MW field which causes
instability of the zero-dc-current state. The Coulomb interaction,
increasing with electron density, is shown to eliminate this effect, and at high
densities (above $10^{7} \mathrm{cm^{-2}}$) the theory presented here yields the dc conductivity behavior which is very
similar to that observed in experiments. We found also that in the high density range, instability
could be triggered by electron heating which restores the applicability of the single-electron
theory.


\begin{thebibliography}{9}

\bibitem{ZudSim-2001} M.A. Zudov, R.R. Du, J.A. Simmons, and J.R. Reno,
Phys. Rev. B {\textbf{64}}, 201311(R) (2001).

\bibitem{ManSme-2002} R. Mani, J.H. Smet, K. von Klitzing, V. Narayanamurti,
W.B. Johnson, and V. Umansky, Nature \textbf{420}, 646 (2002).

\bibitem{ZudDu-2003} M.A. Zudov, R.R. Du, L.N. Pfeiffer, and K.W. West,
Phys. Rev. Lett. \textbf{90}, 046807 (2003).

\bibitem{DurSac-2003} A.C. Durst, S. Sachdev, N. Read, and S.M. Girvin, Phys. Rev.
Lett. \textbf{91}, 086803 (2003).

\bibitem{RyzSur-2003} V. Ryzhii and R. Suris, J. Phys.: Cond. Matt. \textbf{15}, 6855
(2003).


\bibitem{DmiVav-2005} I.A. Dmitriev, M.G. Vavilov, I.L. Aleiner, A.D. Mirlin, and
D.G. Polyakov, Phys. Rev. B \textbf{71}, 115316 (2005).

\bibitem{Mik-2011} S.A. Mikhailov, Phys. Rev. B \textbf{83}, 155303 (2011).

\bibitem{CheShe-2009} A.D. Chepelianskii and D.L. Shepelyansky, Phys. Rev. B
\textbf{80}, 241308(R) (2009).

\bibitem{AndAleMil-2003} A.V. Andreev, I.L. Aleiner, and A.J. Millis, Phys.
Rev. Lett., \textbf{91}, 056803 (2003).

\bibitem{MonKon-book-2004} Yu.P. Monarkha and K. Kono, \textit{Two-Dimensional
Coulomb Liquids and Solids} (Springer-Verlag, Berlin, 2004).

\bibitem{KonKon-2009} D. Konstantinov and K. Kono, Phys. Rev. Lett. \textbf{103},
266808 (2009).

\bibitem{KonKon-2010} D. Konstantinov and K. Kono, Phys. Rev. Lett. \textbf{105},
226801 (2010).

\bibitem{Mon-2011} Yu.P. Monarkha, \textit{Fiz. Nizk. Temp.} \textbf{37}, 108 (2011)
[Low Temp. Phys. \textbf{37}, 90 (2011)]; Yu.P. Monarkha,
\textit{Fiz. Nizk. Temp.} \textbf{37}, 829 (2011) [Low Temp. Phys. \textbf{37}, 655 (2011)].

\bibitem{KonMonKon-2013} D. Konstantinov, Yu.P. Monarkha, and K. Kono, Phys. Rev. Lett.
\textbf{111}, 266802 (2013).

\bibitem{Ryz-1969} V.I. Ryzhii, \textit{Fiz. Tverd. Tela} \textbf{11}, 2577 (1969)
[Sov. Phys. Solid State \textbf{11}, 2078 (1970)].

\bibitem{Par-2004} K. Park, Phys. Rev. B \textbf{69}, 201301 (R) (2004).

\bibitem{LyaPat-2006} I.I. Lyapilin and A.E. Patrakov, Phys. Met.
Metallogr., \textbf{102}, 560 (2006).

\bibitem{DmiMirPol-2012} I. A. Dmitriev, A. D. Mirlin, D. G. Polyakov, and M. A. Zudov,
Rev. Mod. Phys. 84, 1709 (2012).

\bibitem{Mik-2014} S. A. Mikhailov, Phys. Rev. B \textbf{89}, 045410 (2014).

\bibitem{Hus-53} K. Husimi, Prog. of Theor. Phys., \textbf{9}, 238 (1953).

\bibitem{Ryz-2003} V. Ryzhii, Phys. Rev. B \textbf{68}, 193402 (2003).

\bibitem{DykKha-1979} M.I. Dykman and L.S. Khazan, \textit{Zh. Eksp. Teor. Fiz.}
\textbf{77}, 1488 (1979) [Sov. Phys. JETP \textbf{50}, 747 (1979)].

\bibitem{FanDykLea-1997}  C. Fang-Yen, M.I. Dykman, and M.J. Lea, Phys. Rev. B
\textbf{55}, 16272 (1997).

\bibitem{Mon-2012} Yu.P. Monarkha, \textit{Fiz. Nizk. Temp.} \textbf{38}, 579 (2012)
[Low Temp. Phys. \textbf{38}, 451 (2012)].

\bibitem{Mon-2014} Yu.P. Monarkha, \textit{Fiz. Nizk. Temp.} \textbf{40}, 623 (2014)
[Low Temp. Phys. \textbf{40}, 482 (2014)].

\bibitem{PenTreShi-2000} F.C. Penning, O. Tress, H. Bluyssen, E. Teske, M. Seck, P. Wyder,
V.B. Shikin, Phys. Rev. B \textbf{61}, 4530 (2000).

\bibitem{BadAbdKon-2014} A.O. Badrutdinov, L.V. Abdurakhimov, and D. Konstantinov,
Phys. Rev. B \textbf{90}, 075305 (2014).


\end{thebibliography}
\end{document}